\documentclass{article}
\usepackage{amsthm}
\usepackage{amssymb}
\usepackage{amsmath}
\usepackage{amsthm}
\usepackage{amsbsy}

\usepackage{graphics}

\begin{document}
\begin{center}
{\large\bf Brain dynamics through spectral-structural neuronal networks}
\end{center}

\begin{center}
Maricel Agop $^{1,}\dagger$, Alina Gavrilu\c{t} $^{1*,}$, Gabriel Crumpei$^{2,\dagger}$, Mitic\u{a}
Craus$^{3,\dagger}$, and Vlad B\^{i}rlescu $^{4,\dagger}$
\end{center}

\begin{center}
$^{1}\dagger$ Department of Physics, Gheorghe Asachi Technical University of Ia\c{s}i,
Bd. D. Mangeron, no.67, Ia\c{s}i, 700050, Romania, m.agop@yahoo.com\\
$^{1*,}$ Faculty of Mathematics, Al. I. Cuza University, Carol I Bd. 11, Ia\c{s}i,
700506, Romania, gavrilut@uaic.ro\\
$^{2,\dagger}$ Psychiatry, Psychotherapy and Counseling Center Ia\c{s}i, Ia\c{s}i,
700115, Romania, crumpei.gabriel@yahoo.com\\
$^{3,\dagger}\; ^{4,\dagger}$ Technical University Gheorghe Asachi of Ia\c{s}i, Romania, Faculty of
Automatic Control and Computer Engineering, Department of Computer Science
and Engineering 53 A Mangeron Blvd., Ia\c{s}i, RO-700050, Romania E-mails:
craus@cs.tuiasi.ro, v.birlescu@cs.tuiasi.ro
\end{center}
\vspace*{1cm}

{\bf Abstract.} Starting from the morphological-functional assumption of
the fractal brain, a mathematical model is given by activating brain's
non-differentiable dynamics through the\ determinism-nondeterminism
inference of the responsible mechanisms. The postulation of a scale
covariance principle in Schr\"{o}dinger's type representation of the
brain geodesics implies the spectral functionality of the brain dynamics
through mechanisms of tunelling, percolation etc., while in the
hydrodynamical type representation, it implies their structural
functionality through mechanisms of wave schock, solitons type etc. For
external constraints proportional with the states density, the fluctuations
of the brain stationary dynamics activate both the spectral neuronal
networks and the structural ones through a mapping principle of two
distinct classes of cnoidal oscillation modes. The spectral-structural
compatibility of the neuronal networks generates the communication
codes of algebraic type, while the same compatibility on the
solitonic component induces a strange topology (the direct product of the
spectral topology and the structural one) that is responsible of the
quadruple law (for instance, the nucleotide base from the human DNA
structure). Implications in the elucidation of some neuropsychological
mechanisms (memory's location and functioning, dementia etc.) are also
presented.



\section{Introduction}

Many phenomena with complex patterns and structures are widely observed in
brain. These phenomena are some manifestations of a multidisciplinary
paradigm called emergence or complexity. They share a common unifying
principle of dynamic arrays, namely, interconnections of a sufficiently
large number of simple dynamic units can exibit extremely complex and
self-organizing behaviors.

There are many diverse methods of analysis of the dynamics of complex systems, particularly of those addressed to brain. However, due to the fact that they use only differentiability in the study of systems, all of them involve sophisticated and sometimes ambiguous models (Kandel brain's standard models [1]). In our opinion, a way of analysis necessary in the dynamics of complex systems, especially those related to brain, must respect the recent results related to the harmony between morphology and functioning of a system. Thus, if we only concentrate to brain, from structural point of view, the nature is abundant in examples.

The standard approach, by extrapolating neuron performing way to the description of the performance of the whole brain (Kandel [1]), did not produce expected results, because the neuron is part of a network, and from the complex systems' point of view, the properties of the constitutive elements can be recaptured in the properties of the whole system only as emergent.

On the other hand, the tackling from a quantum perspective, as proposed by different authors (Atmanspacher [2,3]), did not allow for specifying of some functional models of the brain, because at every scale the emergent phenomenon generates new properties which can be recaptured as emergent in the scales that follow.

Consequently, the performance of brain as an assembly, from the psychic life point of view, seems to be best approached from the perspective of complex systems, with their potential components, nondifferential and noncausal from the structural point of view, between which there is a chaotic part structured via attractors, and highlighted within a phase space.

From the very beginning we point out that, both from brain structural (morphological) and from functional (processing) points of view, examples are brought in order to substantiate the idea set forth as the starting one, namely that the brain dynamics at every level are dictated by the brain functional-structural coherence.

For example, studying the branching pattern of dendritic trees of retina neurons, Caserta {\it et al.} [4] identify by box counting, fractal shapes with a fractal dimension of aproximatively 1.7, which can be explained by a diffusion limited aggregation model (Witten and Sander [5]). A fractal structure was observed by Kniffki {\it et al.} [6] for the branching dendrite patterns of thalamic neurons in Golgi impregnated specimens. Moreover, as one can easily observe, the entire neuronal network has a fractal structure.

The fractality is also manifest when the functionality of the brain is considered. Such a statement is based on the idea that a great body of laws governing such a functionality at any scale of resolution, prove to be reducible to a power-type law. As a matter of fact, power laws are to be found both in the functionality and in the structure of brain, which substantiates the idea to suggested in the present work, namely that only the structural-functional unity of the brain can lead to a sound explanation of the complex phenomena met at this level (see Werner [7,8]).

The recording of the brain functions highlighted an electric and magnetic activity correlated with certain brain functions. This electromagnetic activity is spectral in nature, and as such assumes a spectral functionality and processing. One can thereby conclude that the morphology and the fractal structure of the brain should be duplicated by a fractal functioning and processing (de Valois \& de Valois [9,10] -- processing of the visual image, B\'{e}k\'{e}si [11] --  processing of proprioceptive sensoriality).

Also, Lowen and Teich [12,13] suggested that the fractal action potential patterning in auditory nerve may be related to fractal activity in the ion channels of the sensory organs feeding into the auditory nerve: that is, the hair cells in the cochlea.

For example, the fractal activity at the site of neural impulse transmission at the neuromuscular junction. The muscular fibers contract nonlinearly due to fractal type mechanisms, which allows for the nonlinear adaptation of muscular contraction to the environmental necessities (sharp transients from rest to maximum function, functional reserve for continuous effort via nonlinear training in time of the muscle fibers).

Our paper sets out to tackling the old problem of brain-mind duality, whereby the description of brain can be approached according to the laws of physics and chemistry, while the mind cannot. We are thus allowed conclude that the two aspects of the duality mind-brain actually embody a structural and functional unity that can be modeled physically and mathematically, and can be analyzed according the the modern scientific paradigms. The psychic life is thus represented by a complex dynamics of exchange between neural and spectral networks.

In the present paper, the brain's dynamics through spectral-structural
neuronal networks are analyzed. Thus, by the determinism-nondeterminism
inference in brain dynamics, we quit either the classical determinism [14,15],
or the quantum nondeterminism [14,16]. Moreover, the fractal type brain [4,12,13,17,18,19,20] is both morphologically and functionally specified  by activating the
non-differentiable type brain dynamics [21-24].

This paper is organized as follows: in Section 2, the mathematical model is
given, the spectral functionality of the brain dynamics is
presented, the structural functionality of the brain dynamics
is provided and the communication codes generation are explained
through the coherence of the spectral-structural functionality. In Section
3, some possible implications of the mathematical model in the decipher of
some neuropsychological mechanisms are presented.

With our approach we finally aim at the solution of the old problem of the duality brain-mind. Our physical-mathematical model offers the vision that both the mind and the brain do form a functional, describable through the dynamics existing between the spectral and neural networks lying at the foundation of the psychic life.

\section{Results}

First, we present the mathematical model, which explains the structure and the functionality of the brain.

The brain is, morphologically, a fractal (see the examples from introduction). Moreover, its own space (the one generated by the brain) is
structurally, a fractal in the most general sense given by Mandelbrot [24].
In such space, the only possible functionalities (which are compatible with
the brain structure) are
achieved on continuous but non-differentiable curves [21-28].

Brain's structural-functional compatibility (structural-functional duality) as a source of the cerebral dynamics at any scale is thus imposed.

Accepting the structural-functional duality of the brain, the trajectory of motion realized on the structural component must be identified with an element from the functional part. If, according to Caserta {\it et al.} [29], we admit that the unharmonic oscillations  of the neurofibrils would be the source of the functional part of the brain, then the curve describing the motion of a neurofibrile is a continuous nondifferentiable curve (see Werner [7,8]). So this motion takes place in a fractal space, the one generated by the fractal structure of the brain, and thus it can be identified with the geodesic of the associated fractal space. At yet another scale, the neuron can be identified with its corresponding geodesic. More generally, the wave is identified with the corpuscle, the motion of the corpuscle in the field of its associated wave being obviously a continuous nondifferentiable curve (fractal curve), whence the idea of geodesic. We shall detail these considerations in what follows:

By "brain dynamics" we understand the application between structural component ("space" variables) and functional component ("time" variables). Due to the fact that the dynamics reflects different levels of application, the time gets in through scale resolution and is denoted by $\delta t$. We reserve the notation $dt$ for the usual time as in the hamiltonian dynamics; $\delta t$ will be defined through a special substitution principle.

Then the following consequences of the brain dynamics emerge, which will be explained in detail through the present paper:

i) Any continuous but non-differentiable curve of the brain dynamics
(brain non-differentiable curve) is explicitly scale
resolution dependent $\delta t$,  i.e., its length tends to infinity when $%
\delta t$ tends to zero;

We mention that, mathematically speaking, a curve is nondifferentiable if it satisfies the Lebesgue theorem, i.e. its length becomes infinite when the scale resolution goes to zero (Mandelbrot [24]). Consequently, in the limit, a curve is as zig-zagged as one can imagine. Thus it exhibits the property of self-similarity in every one of its points, which can be translated into a property of holography (every part reflects the whole). Doesn't this happen in the brain? Of course. This is why, from the smallest scale to the greatest-sized ones, the brain is a whole! [2,3,30];
\medskip

ii) The physics of the brain phenomena is related to the behaviour of a
set of functions during the zoom operation of the scale resolution $\delta t.$ Then, through the
substitution principle, $\delta t$ will be identified with $dt$, i.e., $%
\delta t\equiv dt$ and, consequently, it will be considered as an
independent variable;\medskip

The fractal variables are dynamical variables, depending both on space coordinates and time, as well as on the resolution scale. Then, a difference should be made for instance between the variables describing the dynamics at the nanoscale (induced by neurofibrils), the dynamics at the synapse level, nerve impulse transmission at the dendrite level (for details see Kniffki {\it et. al} [6]). It is the global structural-functional coherence of our brain, which, in our opinion, determines a permanent interdependence between these variables;

iii) The brain dynamics is described through fractal
variables, i.e., functions depending both on the space-time
coordinates and the scale resolution since the
differential time reflection invariance of any  dynamical
variable is broken. Then, in any  point of the
non-differentiable curve, two derivatives of the  variable field $%
Q(t,dt)$ can be defined:
\begin{eqnarray}
\frac{d_{+}Q(t,dt)}{dt} &=&\underset{\Delta t\rightarrow 0_{+}}{\lim }\frac{%
Q(t+\Delta t,\Delta t)-Q(t,\Delta t)}{\Delta t} \\
\frac{d_{-}Q(t,dt)}{dt} &=&\underset{\Delta t\rightarrow 0_{-}}{\lim }\frac{%
Q(t,\Delta t)-Q(t-\Delta t,\Delta t)}{\Delta t}.\nonumber
\end{eqnarray}

The sign $+$ corresponds to the forward process, while
the sign $-$ corresponds to the backwards one;

ii) The differential of the spatial coordinate field $dX^{i}(t,dt)$
is expressed as the sum of the two differentials, one of them being not
scale resolution independent (differential part $d_{\pm
}x^{i}(t))$ and the other one being scale resolution dependent (fractal part $d_{\pm }\xi ^{i}(t))$, i.e.,
\begin{equation}
d_{\pm }X^{i}(t,dt)=d_{\pm }x^{i}(t)+d_{\pm }\xi ^{i}(t,dt);
\end{equation}

v) The non-differentiable part of the brain spatial coordinate field
satisfies the fractal equation:
\begin{equation}
d_{\pm }\xi ^{i}(t,dt)=\lambda _{\pm }^{i}(dt)^{1/D_{F}},
\end{equation}
where $\lambda _{\pm }^{i}$ are constant coefficients through which the
fractalisation type is specified and $D_{F}$ defines the fractal
dimension of the brain non-differentiable curve. Let us note that any
definition (Kolmogorov or Hausdorff-Besicovici fractal dimensions [21-24])
is acceptable for $D_{F}$, but once a certain definition is admitted, it
should be used until the end of the analyzed brain dynamics. Moreover, it
should be considered constant.

In our opinion, the functionality of the cerebral processes implies dynamics on geodesics having various fractal dimensions.
Precisely, for $D_{F}=2$, quantum type functionalities are generated (percolation in living neural networks, tunneling in neurofibrils etc.).
For $D_{F}<2$, correlative type functionalities are generated, while for $D_{F}>2,$ non-correlative type ones can be found (limited or unlimited diffusions-branching pattern of dendritic trees of retina neurons by a diffusion limited aggregation [4-6]).

\medskip

vi) The differential time reflection invariance of any
dynamical variable is recovered by combining the derivates $d_{+}/dt$ and $%
d_{-}/dt$ in the non-differentiable operator
\begin{equation}
\frac{\hat{d}}{dt}=\frac{1}{2}(\frac{d_{+}+d_{-}}{dt})-\frac{i}{2}(\frac{%
d_{+}-d_{-}}{dt}).
\end{equation}

This is a natural result of the in complex through
differentiability [31,32] procedure. Applying now the non-differentiable
operator to the brain spatial coordinate field yields to the brain
complex velocity field
\begin{equation}
\hat{V}^{i}=\frac{\hat{d}X^{i}}{dt}=V_{D}^{i}-V_{F}^{i}
\end{equation}
with
\begin{eqnarray}
V_{D}^{i} &=&\frac{1}{2}(v_{+}^{i}+v_{-}^{i}),V_{F}^{i}=\frac{1}{2}%
(v_{+}^{i}-v_{-}^{i}) \nonumber\\
v_{+}^{i} &=&\frac{d_{+}x^{i}+d_{+}\xi ^{i}}{dt},v_{-}^{i}=\frac{%
d_{-}x^{i}+d_{-}\xi ^{i}}{dt}.
\end{eqnarray}

The real part $V_{D}^{i}$ is differentiable and scale
resolution independent (differentiable velocity field), while
the imaginary one $V_{F}^{i}$ is non-differentiable and
scale resolution dependent (fractal velocity field);\medskip

vii) In the absence of any external  constraint there can be found
an infinite number of non-differentiable curves (brain
geodesics) relating any pair of its points and this is true at
all scales. Then, in the brain fractal space,  the neuron is substituted with the brain geodesics
themselves so that any external constraint
(electroencephalogram, functional MRI etc.) is interpreted as a selection of geodesics by the measuring device. The infinity of brain
geodesics in the bundle, their non-differentiability and the two
values of the derivative imply a generalized statistical fluid like
description (brain non-differentiable fluid). Then the average values of
the brain fluid variables must be considered in the previously mentioned
sense, so the average of $d_{\pm }X^{i}$ is
\begin{equation}
<d_{\pm }X^{i}>\equiv d_{\pm }x^{i}
\end{equation}
with
\begin{equation}
<d_{\pm }\xi ^{i}>=0;
\end{equation}

viii) The brain dynamics can be described through a covariant
derivative, whose explicit form is obtained as follows.

Let us now consider that the non-differentiable curves are
immersed in a $3-$dimensional space (the brain dynamics consciousness is
$3-$dimensional) and that $X^{i}$ are the spatial coordinate field
of a point on the non-differentiable curve. We also
consider a variable field $Q(X^{i},t)$ and the following Taylor
expansion up to the second order
\begin{equation}
d_{\pm }Q(X^{i},t)=\partial _{t}Qdt+\partial _{i}Qd_{\pm }X^{i}+\frac{1}{2}%
\partial _{l}\partial _{k}Qd_{\pm }X^{l}d_{\pm }X^{k}.
\end{equation}

These relations are valid in any point of the space and
more for the points $X^{i}$ on the non-differentiable
curve which we have selected in $(9).$

From here, forward and backward value of $(9)$ become
\begin{equation}
<d_{\pm }Q>=<\partial _{t}Qdt>+<\partial _{i}Qd_{\pm }X^{i}>+\frac{1}{2}%
<\partial _{l}\partial _{k}Qd_{\pm }X^{l}d_{\pm }X^{k}>.
\end{equation}

We supose that the average value of variable field $Q$
and its derivatives coincide with themselves and the differentials
$d_{\pm }X^{i}$ and $dt$ are independent. Therefore, the average of their
products coincides with the product of averages. Consequently, $(10)$
becomes
\begin{equation}
d_{\pm }Q=\partial _{t}Qdt+\partial _{i}Q<d_{\pm }X^{i}>+\frac{1}{2}\partial
_{l}\partial _{k}Q<d_{\pm }X^{l}d_{\pm }X^{k}>.
\end{equation}

Even the average value of $d_{\pm }\xi ^{i}$ is null, for the higher order
of $d_{\pm }\xi ^{i}$, the situation can still be different. Let us focus on
the averages $<d_{\pm }\xi ^{i}d_{\pm }\xi ^{l}>.$ Using $(3)$ we can write
\begin{equation}
<d_{\pm }\xi ^{i}d_{\pm }\xi ^{l}>=\pm \lambda _{\pm }^{i}\lambda _{\pm
}^{l}(dt)^{(2/D_{F})-1}dt,
\end{equation}
where we accepted that the sign $+$ corresponds to $dt>0$ and the the sign $%
- $ corresponds to $dt<0.$

Then $(11)$ takes the form
\begin{equation}
d_{\pm }Q=\partial _{t}Qdt+\partial _{i}Qd_{\pm }x^{i}+\frac{1}{2}\partial
_{i}\partial _{l}Qd_{\pm }x^{i}d_{\pm }x^{l}\pm \frac{1}{2}\partial
_{i}\partial _{l}Q[\lambda _{\pm }^{i}\lambda _{\pm
}^{l}(dt)^{(2/D_{F})-1}dt].
\end{equation}

If we divide by $dt$ and neglect the terms that contain differential factors
(for details see the method from [21-28]) we obtain:
\begin{equation}
\frac{d_{\pm }Q}{dt}=\partial _{t}Q+v_{\pm }^{i}\partial _{i}Q\pm \frac{1}{2}%
\lambda _{\pm }^{i}\lambda _{\pm }^{l}(dt)^{(2/D_{F})-1}\partial
_{i}\partial _{l}Q.
\end{equation}

These relations also allow us to define the operators
\begin{equation}
\frac{d_{\pm }}{dt}=\partial _{t}+v_{\pm }^{i}\partial _{i}\pm \frac{1}{2}%
\lambda _{\pm }^{i}\lambda _{\pm }^{l}(dt)^{(2/D_{F})-1}\partial
_{i}\partial _{l}.
\end{equation}

Under these circumstances, taking into account $(4),(5)$ and $(15)$ let us
calculate $\hat{d}/dt.$ It results
\begin{equation}
\frac{\hat{d}Q}{dt}=\partial _{t}Q+\hat{V}^{i}\partial _{i}Q+\frac{1}{4}%
(dt)^{(2/D_{F})-1}D^{lk}\partial _{l}\partial _{k}Q,
\end{equation}
where
\begin{eqnarray}
D^{lk} &=&d^{lk}-i\overline{d}^{lk} \\
d^{lk} &=&\lambda _{+}^{l}\lambda _{+}^{k}-\lambda _{-}^{l}\lambda _{-}^{k},%
\overline{d}^{lk}=\lambda _{+}^{l}\lambda _{+}^{k}+\lambda _{-}^{l}\lambda
_{-}^{k}.\nonumber
\end{eqnarray}

The relation $(16)$ also allows us to define the covariant
derivative
\begin{equation}
\frac{\hat{d}}{dt}=\partial _{t}+\hat{V}^{i}\partial _{i}+\frac{1}{4}%
(dt)^{(2/D_{F})-1}D^{lk}\partial _{l}\partial _{k}.
\end{equation}

Now, we shall refer to what we call brain's geodesic. 

Let us consider the principle of scale covariance (the
physics laws are invariant we respect to scale transformations)
and postulate that the passage from the differentiable mathematical model to the non-differentiable mathematical model can be implemented
by replacing the standard time derivative $d/dt$ by the non-differentiable
operator $\hat{d}/dt$. Thus, this operator plays the role of the
covariant derivative, namely, it is used to write the fundamental
equations of brain dynamics under the same form as in the classical
(differentiable) case. In these conditions, applying the operator $(18)$ to
the complex velocity field $(5)$, the brain geodesics in the
presence of an external constraint given by the scalar potential $%
U,$ have the following form$:$%
\begin{equation}
\frac{\hat{d}\hat{V}^{i}}{dt}=\partial _{t}\hat{V}^{i}+\hat{V}^{l}\partial
_{l}\hat{V}^{i}+\frac{1}{4}(dt)^{(2/D_{F})-1}D^{lk}\partial _{l}\partial _{k}%
\hat{V}^{i}=-\partial ^{i}U.
\end{equation}

This means that the local acceleration $\partial _{t}\hat{V}^{i}$,
the convection $\hat{V}^{l}\partial _{l}\hat{V}^{i}$, the
dissipation $D^{lk}\partial _{l}\partial _{k}\hat{V}^{i}$ and the
forces induced by the external constraints $\partial ^{i}U$ make
their balance in any point of the brain non-differentiable
curve. Moreover, the presence of the complex coefficient of
viscosity type $\frac{1}{4}(dt)^{(2/D_{F})-1}D^{lk}$ specifies that the
neuronal medium is a rheological medium, so it has memory, as a
datum, by his own structure.

If the fractalisation is achieved by Markov type stochastic
processes [21,22,24], then
\begin{equation}
\lambda _{+}^{i}\lambda _{+}^{l}=\lambda _{-}^{i}\lambda _{-}^{l}=2\lambda
\delta ^{il},
\end{equation}
where $\delta ^{il}$ is Kronecker's symbol with the property
\begin{equation*}
\delta ^{il}=
\begin{cases}
1,&i=l \\
0,&i\neq l.
\end{cases}
\end{equation*}
In these conditions, the equation of brain geodesics takes the simple form
\begin{equation}
\frac{\hat{d}\hat{V}^{i}}{dt}=\partial _{t}\hat{V}^{i}+\hat{V}^{l}\partial
_{l}\hat{V}^{i}-i\lambda (dt)^{(2/D_{F})-1}\partial ^{l}\partial _{l}\hat{V}%
^{i}=-\partial ^{i}U
\end{equation}
or more, by separating the  motions on  differential and
fractal scale resolutions,
\begin{eqnarray}
\frac{\hat{d}V_{D}^{i}}{dt} &=&\partial _{t}V_{D}^{i}+V_{D}^{l}\partial
_{l}V_{D}^{i}-[V_{F}^{l}-\lambda (dt)^{(2/D_{F})-1}\partial ^{l}]\partial
_{l}V_{F}^{i}=-\partial ^{i}U \\
\frac{\hat{d}V_{F}^{i}}{dt} &=&\partial _{t}V_{F}^{i}+V_{D}^{l}\partial
_{l}V_{F}^{i}+[V_{F}^{l}-\lambda (dt)^{(2/D_{F})-1}\partial ^{l}]\partial
_{l}V_{D}^{i}=0\nonumber.
\end{eqnarray}

In what follows, we discuss the spectral functionality through the brain's geodesics of Schr\"{o}dinger type.

For irrotational motions
\begin{equation}
\varepsilon _{ikl}\partial ^{k}\hat{V}^{l}=0,
\end{equation}
where $\varepsilon _{ikl}$ is the L\'{e}vy-Civita pseudo-tensor. We choose $\hat{V}%
^{i} $ in the form which makes this definition an identity
\begin{equation}
\hat{V}^{i}=-2i\lambda (dt)^{(2/D_{F})-1}\partial ^{i}\ln \Psi ,
\end{equation}
where for the moment $\ln \Psi $ defines the scalar potential of
the complex velocity field.

Substituting $(24)$ in $(21)$ we obtain
\begin{eqnarray}
\frac{\hat{d}\hat{V}^{i}}{dt} &=&-2i\lambda (dt)^{(2/D_{F})-1}\{\partial
_{t}\partial ^{i}\ln \Psi -i[2\lambda (dt)^{(2/D_{F})-1}(\partial ^{l}\ln
\Psi \partial _{l})\cdot \partial ^{i}\ln \Psi + \\&&
+\lambda (dt)^{(2/D_{F})-1}\partial ^{l}\partial _{l}\partial ^{i}\ln \Psi
]\} =-\partial ^{i}U\nonumber.
\end{eqnarray}

Using the identities
\begin{eqnarray}
\partial ^{l}\partial _{l}\ln \Psi +\partial _{i}\ln \Psi \partial ^{i}\ln
\Psi &=&\frac{\partial _{l}\partial ^{l}\Psi }{\Psi } \\
\partial ^{i}(\frac{\partial ^{l}\partial _{l}\Psi }{\Psi }) &=&2(\partial
^{l}\ln \Psi \partial _{l})\partial ^{i}\ln \Psi +\partial ^{l}\partial
_{l}\partial ^{i}\ln \Psi \nonumber
\end{eqnarray}
the equation $(25)$ becomes
\begin{equation}
\frac{\hat{d}\hat{V}^{i}}{dt}=-2i\lambda (dt)^{(2/D_{F})-1}\partial
^{i}[\partial _{t}\ln \Psi -2i\lambda (dt)^{(2/D_{F})-1}\frac{\partial
^{l}\partial _{l}\Psi }{\Psi }]=-\partial ^{i}U.
\end{equation}

This equation can be integrated up to an arbitrary phase factor, which may
be set to zero by a suitable choice of phase of $\Psi $ and this yields:
\begin{equation}
\lambda ^{2}(dt)^{(4/D_{F})-2}\partial ^{l}\partial _{l}\Psi +i\lambda
(dt)^{(2/D_{F})-1}\partial _{t}\Psi -\frac{U}{2}\Psi =0.
\end{equation}

The relation $(28)$ is a Schr\"{o}dinger type equation ("brain geodesics
of Schr\"{o}dinger type") and it implies the following:

i) According to [33], $\Psi $ is a wave function and it has a
direct physical signification only through $|\Psi |^{2}$ as probability
density ( probability density);\medskip

ii) The unpredictable character of the brain dynamics is specified
through the wave properties of the neuronal medium (or, neuronal
network). In this way, brain's spectral type functionality is
provided;\medskip

iii) The mechanisms that are responsible of brain's spectral functionality
are of quantum type only when they are extrapolated for various
scale resolutions (such as tunneling, percolation, entanglement states - see Werner [7] etc.).

Now, we refer to the structural functionality through brain geodesics of hydrodynamic type.

If $\Psi =\sqrt{\rho }\exp (iS)$ with $\sqrt{\rho }$ the amplitude and $S$
the phase of $\Psi $, the  complex velocity field $(5)$ takes the
explicit form
\begin{eqnarray}
\hat{V}^{i} &=&2i\lambda (dt)^{(2/D_{F})-1}\partial ^{i}\ln \Psi \nonumber\\
V_{D}^{i} &=&2i\lambda (dt)^{(2/D_{F})-1}\partial ^{i}S \\
V_{F}^{i} &=&2i\lambda (dt)^{(2/D_{F})-1}\partial ^{i}\ln \rho \nonumber.
\end{eqnarray}

Substituting $(29)$ into $(19)$ and separating the real and imaginary parts,
up to an arbitrary phase factor which may be set to zero by a suitable
choice of the phase of $\Psi $, we obtain:
\begin{eqnarray}
\partial _{t}V_{D}^{i}+(V_{D}^{l}\partial _{l})V_{D}^{i} &=&-\partial
^{i}(Q+U) \\
\partial _{t}\rho +\partial ^{i}(\rho V_{D}^{i}) &=&0
\end{eqnarray}
with $Q$ the  specific non-differentiable potential
\begin{equation}
Q=-2\lambda (dt)^{(4/D_{F})-2}\frac{\partial ^{l}\partial _{l}\sqrt{\rho }}{%
\sqrt{\rho }}=-\frac{V_{F}^{l}V_{Fl}}{2}-\lambda (dt)^{(2/D_{F})-1}\partial
_{i}V_{F}^{i}.
\end{equation}

Equation $(30)$ represents the  specific momentum conservation law,
while equation $(31)$ represents the  states density conservation
law. Equations $(30)$ and $(31)$ define the  fractal hydrodynamic
model and imply the following:

i) Any neuronal cell is in a permanent interaction with a fractal medium,
identified with the neuronal network through the  specific
non-differentiable potential $(32)$. The physics fractal medium is prone to
computational properties [34];\medskip

ii) The neuronal network can be identified with a fractal fluid (non-differentiable fluid), whose dynamics is described by the fractal hydrodynamical model;\medskip

iii) The fractal velocity field $V_{F}^{i}$ does not represent
actual motion, but contributes to the transfer of the
specific momentum and to the brain energy focus, thus confering spectral representability to brain functioning through we what call {\it neuronal network.} This may be seen
clearly from the absence of $V_{F}^{i}$ from the states density
conservation law and from its role in the variational principle
[21,22];\medskip

iv) Any interpretation of $Q$ should take cognizance of the self nature
of the specific momentum transfer. While the brain energy is
stored in the form of the  mass motion and  potential
energy, some is available elsewhere and only the total is conserved. It is
the conservation of the energy and the specific momentum
that ensures reversibility and the existence of
eigenstates, but denies a L\'{e}vy type motion [35] of
brain interaction with an external medium;\medskip

v) The predictable character of the brain activity is specified
through the corpuscular properties of the neuronal network. Thus, the brain's
corpuscular type functionality is provided;\medskip

vi) The mechanisms that are responsible of brain's structural functionality
are of a hydrodynamical type, but when they are extrapolated for various
 scale resolutions (shock waves, solitons [36] etc.);\medskip

vii) The specific non-differentiable potential coordinates the
transitory functionality (the spectral-structural functionality).

Next, we shall present several considerations about the generation of the communication languages. 

Both functionalities (either the one which is responsible of the brain
activity unpredictable character, or the other one which is responsible of
the brain activity predictable character) act simultaneously. By their
interconditioning there result either brain coherence or brain
incoherence. Indeed, let us admit that both in the brain representation
of Schr\"{o}dinger type and in the brain representation of hydrodynamical
fractal type, the external constraint is proportional with the states density, i.e., $U=2a|\Psi |^{2}=2a\rho $, with $a=$ const.
Then for the stationay case ($\partial _{t}\Psi =0$ and $\partial _{t}\rho
=0,V_{D}^{i}=\partial ^{i}S=0)$ the equation $(28)$ becomes:
\begin{equation}
\lambda ^{2}(dt)^{(4/D_{F})-2}\partial ^{l}\partial _{l}\Psi +E_{\Psi }\Psi
-a|\Psi |^{2}\Psi =0,
\end{equation}
while the equations $(30)-(32)$ get by integration the form:
\begin{equation}
Q+U=-2\lambda ^{2}(dt)^{(4/D_{F})-2}\frac{\partial ^{l}\partial _{l}\sqrt{%
\rho }}{\sqrt{\rho }}+2a\rho =2E_{\rho }
\end{equation}
with $E_{\Psi }$ and $E_{\rho }$ constants having specific
energies significances.

By the substitutions:
\begin{eqnarray}
\frac{(E_{\Psi })^{1/2}}{\lambda k_{\Psi }}(dt)^{1-(2/D_{F})}(k_{\Psi
x}x+k_{\Psi y}y+k_{\Psi z}z) &=&\xi _{\Psi } \nonumber\\
\frac{(E_{\rho })^{1/2}}{\lambda k_{\rho }}(dt)^{1-(2/D_{F})}(k_{\rho
x}x+k_{\rho y}y+k_{\rho z}z) &=&\xi _{\rho } \nonumber\\
\Psi &=&(\frac{E_{\Psi }}{a})^{1/2}f,\sqrt{\rho }=(\frac{E_{\rho }}{a}%
)^{1/2}h \\
\mathbf{k}_{\Psi }^{2} &=&k_{\Psi x}^{2}+k_{\Psi y}^{2}+k_{\Psi z}^{2},%
\mathbf{k}_{\rho }^{2}=k_{\rho x}^{2}+k_{\rho y}^{2}+k_{\rho z}^{2}\nonumber,
\end{eqnarray}
$(33)$ and $(34)$ reduce to the equations of Ginzburg-Landau type
[36]:
\begin{equation}
\frac{d^{2}f}{d\xi _{\Psi }^{2}}=f^{3}-f
\end{equation}
for spectral functionality, respectively
\begin{equation}
\frac{d^{2}h}{d\xi _{\rho }^{2}}=h^{3}-h,
\end{equation}
for structural functionality, where $\mathbf{k}_{\Psi }$ and $\mathbf{k}%
_{\rho }$ are the brain wave vectors.

Using the methodology from [25], these previous equations admit
\textit{either} the infinite energy solutions
\begin{eqnarray}
|\Psi |^{2} &=&\frac{2s_{\Psi }^{2}}{1+s_{\Psi }^{2}}sn^{2}(\frac{\xi _{\Psi
}-\xi _{\Psi _{0}}}{\sqrt{2}};s_{\Psi }) \\
\rho  &=&\frac{2s_{\rho }^{2}}{1+s_{\rho }^{2}}sn^{2}(\frac{\xi _{\rho }-\xi
_{\rho _{0}}}{\sqrt{2}};s_{\rho })
\end{eqnarray}
where $sn$ are Jacobi's elliptic functions of modules $s_{\Psi }$ and $%
s_{\rho }$ [37]
\begin{equation}
s_{\Psi }=\frac{1-(1-2c_{\Psi })^{1/2}}{1+(1-2c_{\Psi })^{1/2}},s_{\rho }=%
\frac{1-(1-2c_{\rho })^{1/2}}{1+(1-2c_{\rho })^{1/2}}
\end{equation}
with $\xi _{\Psi _{0}},\xi _{\rho _{0}},c_{\Psi },c_{\rho }$ integration
constants, \textit{or} the finite energy solutions (kink
solutions [38])
\begin{eqnarray*}
|\Psi |^{2} &=&\tanh (\frac{\xi _{\Psi }-\xi _{\Psi _{0}}}{\sqrt{2}}) \\
\rho  &=&\tanh (\frac{\xi _{\rho }-\xi _{\rho _{0}}}{\sqrt{2}})
\end{eqnarray*}
obtained through the degeneration of the elliptic functions $sn$ in the
modules $s_{\Psi }$ and $s_{\rho }$, i.e.,
\begin{eqnarray}
s_{\Psi } &\rightarrow &1\text{ for }c_{\Psi }\rightarrow 1/2 \\
s_{\rho } &\rightarrow &1\text{ for }c_{\rho }\rightarrow 1/2.\nonumber
\end{eqnarray}

Now, some conclusions are obvious:

i) Both the probability density fluctuations
\begin{equation}
\delta |\Psi |^{2}=1-\frac{1+s_{\Psi }^{2}}{2s_{\Psi }^{2}}|\Psi
|^{2}=cn^{2}(\frac{\xi _{\Psi }-\xi _{\Psi _{0}}}{\sqrt{2}};s_{\Psi })
\end{equation}
and the states density ones
\begin{equation}
\delta \rho =1-\frac{1+s_{\rho }^{2}}{2s_{\rho }^{2}}\rho =cn^{2}(\frac{\xi
_{\rho }-\xi _{\rho _{0}}}{\sqrt{2}};s_{\rho })
\end{equation}
obtained based on the infinite energy solutions express oneselves
through the cnoidal oscillations modes of the spatial coordinates fields $%
\xi _{\Psi }$ and $\xi _{\rho }$, where $cn$ are Jacobi's elliptic functions
of modules $s_{\Psi }$ and $s_{\rho }$ [37];\medskip

ii) Generally speaking, as it also results from [36], the
cnoidal oscillation modes are equivalent to one-dimensional Toda type
lattices of nonlinear oscillators [39,40]. Moreover, according to
[14-16], their mapping implies Toda type neuronal networks. That is
why, based on the above results, we shall be able to define two Toda type
neuronal networks, one of them being specific to the spectral functionality
(and which will be called the {\it spectral neuronal network}) and the other
one being specific to the structural functionality (and which will be called
the {\it structural neuronal network});\medskip

iii) Since both functionalities, the spectral one and the structural one,
define the same physical object, these imply the identities:
\begin{equation}
\xi _{\Psi }\equiv \xi _{\rho },E_{\Psi }\equiv E_{\rho },\mathbf{k}_{\Psi
}\equiv \mathbf{k}_{\rho },|\Psi |^{2}\equiv \rho .
\end{equation}

Consequently, the probability density fluctuations are identified
with the states density fluctuations, which specifies the brain
coherence (or brain compatibility) of the two neuronal networks (the
spectral one and the structural one). Such a situation implies
mathematically the functionality of the elliptic functions equivalence
theorem [37] and, in consequence, it implies the existence of certain
algebraic relations among the variables which define brain
dynamics on the two neuronal networks, particularly,
\begin{equation}
\delta |\Psi |^{2}=F(\delta \rho ).
\end{equation}

By the algebraic relations, self-structuring, communication codes
(languages) of the physical object result;\medskip

iv) According to  [27,28], the finite
 energy solutions (see $(41)$ and $(42)$) can be also obtained by
field theories with spontaneous symmetry breaking, which also implies
strange topologies [28,41]. In any of these topologies one can always
define through the associated topological charges, two distinct stable
physical states. Now, according to our results, we shall have on one hand a
topology specific to the spectral functionality (called spectral
topology) which will define the spectral states through the spectral
topological charge and, on the other hand, the topology specific to the
structural functionality (called structural topology) which will define
the structural states through the structural topological charge.
Moreover, due to their physical object status, the two topologies, the
spectral one and the structural one act simultaneously, influencing
one each other. Practically speaking, we have a unique topology which
encompasses both of them in the form of their direct product [41]. This has
as a consequence the existence of four distinct stable states. In our
opinion, these states could be associated with the nucleotide base
from the human DNA structure [42].

\section{Discussion}

In this section, we talk about possible implications of the mathematical model in the decipher
of some neuropsychological mechanisms.

The implementattion of the functional structure of complex systems to psychic life, can explain a series of classical concepts circulated during te last century. Thus the unconscious from psychoanalysis can be associated with the unpredictable, noncausal and potential part from the structure of the complex system, while the conscience, as well as the unconscious behavioral patterns (superego of psychoanalysis) can be associated with the structured, causal and deterministic part of the complex system. Between the two parts, there exist a pemanent dynamics through attractors, descriptible in the phase space. The chaoticity existing between the two components is absolutely necessary for the functioning of the brain. When it is affected by repetitive cycles (epilepsy crisis) the conscience is blurred.

This new representation on brain's functioning leads to new conclusions concerning different mental processes that have not been  fully
understood yet:

i) Consciousness could be the dynamics result between the two networks: the
spectral neuronal network and the structural one. For instance, the
anesthetic techniques block the structural network. When this structural
network becomes again functional, it recovers its dynamics through the
multifocal coherence phenomenon with the spectral network (where the memory,
the core personality can be found, as detailed in the following). The same
thing happens in epileptic crisis, in concussions, electroshocks etc., the
structural network being unable to achieve coherent dynamics with the
spectral network;\medskip

ii) In the structure of the brain (as a physical object), the memory may be
located in the spectral neuronal network, whose spectral, and thus fractal
character has all the properties that are necessary for the information storage.
The memory means coherence achievement among certain structures of the
structural neuronal network and the spectral one, where those information
have been memorized;\medskip

iii) Memory localization could give clues on how personality is structured.
Classically, the personality has two components: temperament and character. The
temperament is constituted of behavioral and information processing patterns
originating from the genetic setting (and which are organized in the
structural network). The second component, the character, represents the
programs built from the individual relationship with the external medium
(education, experiences, cultural environment, analyzers' setting etc.). It
is organized in the spectral network, representing information, behavioral
and information processing patterns, that are set in programs resulting from
the system and the external medium dynamics. The genetic patterns found in
the structural neuronal network give stability and the programs built in the
spectral neuronal network are adapted to the environment, in a specific form
given by the dynamics with the genetic patterns from the structural network.
In this way, personality has stability via some of its components, but it
also has specificity and adaptability;\medskip

iv) It seems that the potentiality Chomski [43] was talking about, related
to every child's ability to learn the language or the languages to which he is
exposed, is related to the spectral neuronal network which gives the memory
space, while the structural network represented by Wernicke and Broca center
(of speech understanding and speech expression, built by patterns
transmitted at the genetic information level) offers the language processing
structure;\medskip

v) In the context of new discoveries about mirror neurons, our model concerning
psyche's functionality and structuring could give explanations about mirror
neurons' functioning and their integration in the psychological functioning
in general. So far, experimental data emphasizes only the elements from the
structural neuronal network (excited neurons, highlighted by electrodes
implanted or brain areas highlighted by fMRI). Accepting the spectral
network could explain complex phenomena, concepts, feelings that could not
be generated only by the activity of several neurons, but by complex
processing that could take place only in the spectral neuronal network. It
might even be possible that the neurons excitation is achieved through the
spectral network, where the information originates through interpersonal
communications spectral vibrational ways. It could be thus explained a
series of controversies about mimetic learning, empathy, mind theory,
language etc.;\medskip

vi) In neuropathology, our model could also generate new conclusions
concerning both mental and neuropsychological illness. For instance, in
vascular dementia, the blood deficiency affects on one hand the neurons (the
structural neuronal network) and on the other hand, it influences the
dynamics between the two networks, while in Alzheimer dementia, the dynamics
between the two networks is primarily affected, with the impossibility to
access the information stored, with the damage of the spatial-temporal
orientation, but also of the behavior and even of the entire personality
(see the considerations from ii)). Certain somatic trauma cases when the
phantom limb sensation manifests itself (Ramachandran [44]) could have
an explanation by the model we conceived; the structural network can be
inhibited or destroyed by the respective limb or organ, its representation
remaining in the spectral neuronal network, generating the painful and
contracted phantom limb symptoms and allowing the alleviating and curing
through a suggestion and autosuggestion mechanism (the mirror box
technique);\medskip

vii) The neuroplasticity phenomenon related to brain's adaptive capacity
would be more understandable if, causally, according to our model, the
neurons and the neuronal connections development would achieve by the
dynamics between the two networks, based on the patterns developed in the
spectral side from the reaction with the environment.

\section{Conclusion}

With our approach we finally aim at the solution of the old problem of the duality brain-mind. Our physical-mathematical model offers the vision that both the mind and the brain do form a functional, describable through the dynamics existing between the spectral and neural networks lying at the foundation of the psychic life.

We made obviously that the brain dynamics can be represented by a Bohmian mechanics (Bohm [30]) exactly the way the regular Schr\"{o}dinger wave mechanics is represented. Only here there are now hidden parameters. Their place is explicitly filled by neuronal medium.

The main conclusions of the present paper are the following:

i) The morphological-functional assumption of the fractal brain (in the most
general sense of this concept described by Mandelbrot [24]) induces a mathematical
model by activating brain's non-differentiable dynamics. Thus, one gives
up either on the classical determinism, or on the quantum nondeterminism
through the determinism-nondeterminism inference of the responsible
mechanisms;\medskip

ii) Through the mathematical model, the scale covariance principle
induces brain geodesics in the velocity representation, situation in
which any neuron from the network is substituted with the brain geodesics
themselves and, moreover, any external constraint
(electroencephalogram, functional MRI etc.) is interpreted as a selection of
the brain geodesics by the measuring device;\medskip

iii) Through the Schr\"{o}dinger type representation of the brain
geodesics, the spectral (wave) functionality of the brain dynamics is
explicited, which implies specific mechanisms of tunneling type, enteglement
states type etc., while through the hydrodynamical representation, the
structural (corpuscular) functionality of the brain dynamics is explicited,
which induces specific mechanisms of shock waves type, solitons type
etc.;\medskip

iv) The inference of these two representations sets forth the transitory
(spectral-structural) functionality, which is controlled by the
specific non-differentiable potential;\medskip

v) If the external constraint is proportional with the
states density, then the stationary dynamics through their
fluctuations activate both the spectral neuronal network and the
structural neuronal network. These classes of neuronal networks result
by the mapping of certain classes of one-dimensional Toda type
networks. Usually, the one-dimensional Toda type networks are associated
with certain cnoidal oscillation modes that, in our situation, can be
specific either to the spectral character, or to the corpuscular
one;\medskip

vi) The structural-functional compatibility (coherence) of these two
classes generates classes of algebraic type communication
codes;\medskip

vii) The spectral-structural compatibility of the neuronal networks only
on the solitonic component simultaneously activates the functionality of
a strange topology (the direct product of the spectral topology and the
structural topology), which induces four distinct physical states through
the associated topological charges. Thus, the quadruple logic
elements are generated. In our opinion, such logic could be associated
with the nucleotide base from the human DNA structure;\medskip

viiii) Possible implications of the mathematical model in the
elucidation of some neuropsychological mechanisms (for instance,
memory's location and functioning, brain trauma consequences, dementia etc.)
are presented.\bigskip \bigskip

{\bf Author Contributions}

In this paper, Maricel Agop provided the original idea and constructed its
framework. Together with Alina Gavrilu\c{t}, the detailed calculation was
conducted, and they were responsible for drafting and revising the whole
paper. The implications and correspondences in neuroscience activity of this
model were given by Gabriel Crumpei. The role in communication codes was
highlighted by Mitic\u{a} Craus and Vlad B\^{i}rlescu. All authors have read
and approved the final manuscript.


\bigskip
{\bf Acknowledgments.}
The authors are indebted to Nicolae Mazilu for valuable discussions and remarks.

\end{document}